\documentstyle[12pt]{article}

\catcode`@=11
\font\twlveufm=eufm10 at 12pt
\font\teneufm=eufm10 
\font\seveneufm=eufm7
\font\twlvmsb=msbm10 at 12pt
\font\tenmsb=msbm10

\def\mset#1#2#3#4{\textfont#1#2\scriptfont#1#3\scriptscriptfont#1#4}
\newfam\eufmfam

\mset\eufmfam\twlveufm\teneufm\seveneufm

\def\frak#1{{\fam\eufmfam\relax#1}}

\newfam\msbfam
\mset\msbfam\twlvmsb\tenmsb\tenmsb

\def\newsymbol#1#2#3#4#5{\let\next@\relax
 \ifnum#2=\@ne\let\next@\msafam@\else
 \ifnum#2=\tw@\let\next@\msbfam@\fi\fi
 \mathchardef#1="#3\next@#4#5}

\def\newsymbol#1#2#3#4#5{\let\next@\relax
 \ifnum#2=\@ne\let\next@\msafam@\else
 \ifnum#2=\tw@\let\next@\msbfam@\fi\fi
 \mathchardef#1="#3\next@#4#5}

\def\Bbb#1{{\fam\msbfam\relax#1}}
\def\hexnumber@#1{\ifcase#1 0\or 1\or 2\or 3\or 4\or 5\or 6\or 7\or 8\or
 9\or A\or B\or C\or D\or E\or F\fi}

\edef\msbfam@{\hexnumber@\msbfam}
\newsymbol\ltimes 226E

\catcode`@=12

\def\square#1{{
\dimen0=#1pt
\dimen1= #1pt
\advance \dimen1 by -.5pt
\vbox
{ \hrule width \dimen0
  \hbox
  {\vrule\hskip\dimen1\vbox to \dimen1{}\vrule}
  \hrule width \dimen0
}}}

\def\strutdepth{\dp\strutbox}
\def\margnote#1{\strut\vadjust{\kern-\strutdepth\specialstar{#1}}}
\def\specialstar#1{\vtop to \strutdepth{
   \baselineskip\strutdepth
   \vss\llap{\footnotesize #1 }\null}}

\def\qed{\ifvmode\removelastskip\penalty100\fi
\hbox{}\nobreak\hfill\nobreak\square{8}}

\newtheorem{theorem}{Theorem}[section]
\newtheorem{corollary}[theorem]{Corollary}
\newtheorem{lemma}[theorem]{Lemma}
\newtheorem{proposition}[theorem]{Proposition}
\newtheorem{conjecture}[theorem]{Conjecture}
\newtheorem{definition}[theorem]{Definition}
\newenvironment{proof}{\par\noindent{\it Proof: }}{\qed\medbreak}
\renewenvironment{abstract}{\quote \noindent{\sc Abstract.}}{\endquote}

\def\E{{\rm e}}
\def\I{{\rm i}}

\def\lhooksym#1{{
\dimen0=#1pt
\dimen1=#1pt
\divide\dimen1 by 2
\dimen2=\dimen1
\advance\dimen2 by .5pt
\vbox
{ \hbox{\hskip \dimen1\vbox to \dimen0{}\vrule}
  \hrule width \dimen2
}}}
\def\setbrak#1{\left\{#1\right\}}
\def\mone{{\hbox{-}1}}
\def\supl{^{\scriptscriptstyle \rm L}}
\def\ssr{{\scriptscriptstyle \rm R}}
\def\supr{^\ssr}

\def\supp{^{\pi}}

\def\where{\quad\mbox{ where }}

\def\ap{a\supp}

\def\Cp{C\supp}
\def\hpr{(\hal^\perp)\supr}
\def\hr{{\hal\supr}}
\def\hp{{\hal^\perp}}
\def\Cl{{C\supl}}

\def\fal{{\frak f}}
\def\gal{{\frak g}}
\def\hal{{\frak h}}

\def\aal{{\frak a}}
\def\Gsp{{\bf G}}
\def\Hsp{{\bf H}}
\def\Msp{{\bf M}}
\def\Usp{{\bf U}}
\def\Ssp{{\bf S}}

\def\Rsp{{\bf R}}
\def\Fsp{{\bf F}}

\def\rT{{\rm T}}

\def\reals{{\Bbb R}}

\def\cmplx{{\Bbb C}}
\def\Pb{{\Bbb P}}
\def\natnums{{\Bbb N}}
\def\Hb{{\Bbb H}}

\def\GL{{\rm GL}}

\def\slf{{\frak{sl}}}
\def\gl{{\frak{ gl}}}
\def\cH{{\cal H}}
\def\cL{{\cal L}}

\def\cD{{\cal D}}
\def\rank{\mathop{\rm rank}\nolimits}

\def\Ad{\mathop{\rm Ad}\nolimits}

\def\Sym{\mathop{\rm Sym}\nolimits}

\def\twomat#1{\left(\begin{array}{rr}#1\end{array}\right)}
\def\mat#1#2{\left(\begin{array}{#1}#2\end{array}\right)}

\def\lparen{\left(}
\def\rparen{\right)}
\def\lp{\lparen}
\def\rp{\rparen}

\def\tnabla{\tilde{\nabla}}

\def\al{\alpha}
\def\alr{{\alpha\supr}}
\def\cfty{{\cal C}^{\raise1pt\hbox{$\scriptscriptstyle \infty$}}\!}
\def\comega{{\cal C}^{\raise1pt\hbox{$\scriptscriptstyle \omega$}}\!}
\def\beq{\begin{equation}}
\def\eeq{\end{equation}}

\title{Imprimitively generated Lie-algebraic Hamiltonians and
  separation of variables.}
\author{
Robert Milson\\
McGill University}
\date{June 1998}
\begin{document}
%\bibliographystyle{plain}
%\include{titlepage}
%\include{section1}
%\include{section2}
%\include{section3}
%\include{section4}
%\include{section5}

%\bibliography{qes}

\begin{titlepage}
\maketitle
\thispagestyle{empty}
\begin{abstract}
  Turbiner's conjecture posits that a Lie-algebraic Hamiltonian
  operator whose domain is a subset of the Euclidean plane admits a
  separation of variables.  A proof of this conjecture is given in
  those cases where the generating Lie-algebra acts imprimitively.
  The general form of the conjecture is false.  A counter-example is
  given based on the trigonometric Olshanetsky-Perelomov potential
  corresponding to the $A_2$ root system.
\end{abstract}


\bigskip
\noindent This research supported by an NSERC, Canada post-doctoral fellowship.

\end{titlepage}

%%% Local Variables: 
%%% mode: latex
%%% TeX-master: "paper3"
%%% End: 
\section{Introduction}
\subsection{Motivation}

There are profound connections between quantum mechanics and the
theory of Lie algebras, and their representations.  Typically, one
looks for a way to relate a given Hamiltonian operator to a
finite-dimensional Lie-algebra, and then uses information about the
algebra's representations to solve the corresponding spectral problem.
This general philosophy manifests itself in a number of distinct
approaches \cite{dgroups} \cite{paldus} \cite{grpscat}.  The context
for the present article is the application of Lie theory to the study
of quasi-exactly solvable spectral problems.  In order to motivate the
questions dealt with here it will be useful to briefly review the
relevant background.

A spectral problem is called quasi-exactly solvable (Q.E.S. for short)
if there exists a method for explicitly obtaining eigenvalues and
eigenvectors for a finite subset of an otherwise infinite spectrum
\cite{ushveridze}.  Typically, quasi-exact solvability amounts to the
existence of an explicitly describable basis of a finite-dimensional
invariant subspace.  The Q.E.S. problems related to 1-dimensional
non-relativistic quantum mechanics are profoundly related to the
following realization of $\slf_2\reals$ by first-order differential
operators \cite{tusl2} \cite{KO:sl2}:
$$T_1=\partial_z,\quad T_2=2z\partial_z-n,\quad T_3=z^2\partial_z-n
z,\qquad n\in \natnums.$$
The crucial fact regarding these operators
is that the vector space of polynomials of degree $n$ or less is an
invariant subspace; indeed this subspace realizes the $n+1$
dimensional, irreducible representation of $\slf_2\reals$.

It therefore stands to reason that every differential operator
generated by the $T_i$'s will admit the same finite-dimensional
invariant subspace.  Consider, for instance, the operator given by
$$-\cH_0 = \frac{1}{4}\, (T_1 T_2+T_2 T_1) + 16 b \, T_3 + 8c\, T_2 +
\frac{n}{2}\, T_1,
$$
where $b,c$ are real parameters.  A change of coordinates,
$z=x^2/4$, and a gauge transformation,
$\cH=\mu\cdot\cH_0\cdot\mu^\mone$, where
$$\mu=\exp\lp {b\over4} x^4 + {c\over2} x^2\rp,$$
yields the Hamiltonian
for the sextic anharmonic potential \cite{znojil} \cite{magyar}:
$$\cH = -\partial_{xx} + b^2 x^6 + 2bc x^4 + (c^2 + b(3+4n)) x^2+c(1+2n).$$
This Hamiltonian operator is quasi-exactly solvable because the
method of construction guarantees that the vector space spanned by $\{
\mu, \mu x^2, \ldots , \mu x^{2n}\}$ is an invariant subspace of the
operator.

The above construction of a Q.E.S. Hamiltonian has the following
generalization to higher dimensions \cite{ShTu:mdim} \cite{GKO:mdim}.
One starts with a realization of a finite-dimensional Lie algebra
$\gal$ by first order differential operators
$$T_a=V_a+\lambda_1\eta_{1a}+\ldots + \lambda_k\eta_{ka},\quad
a=1,\ldots, \dim(\gal),$$
where the $V_a$'s are a realization of
$\gal$ by vector fields, where the $\eta_{ia}$ are functions, and
where the $\lambda_i$ are real parameters such that for certain values
there exists a finite-dimensional invariant subspace of functions,
$W_\lambda$.

The general second-order order, Lie algebraic operator is given by:
\begin{equation}
\label{laoperator.eqn}
\cH_0 = \sum_{ab} C^{ab} T_a T_b + \sum_a
B^a T_a, 
\end{equation}
where the $C^{ab}$ and $B^a$ are real numbers with
$C^{ab}=C^{ba}$.  In the rest of this article operators like $\cH_0$,
i.e. operators that can be generated by a finite-dimensional Lie
algebra of first-order differential operators, will be called
Lie-algebraic.  A Lie algebraic Hamiltonian/Schr\"odinger operator,
$\cH$, is a formally self-adjoint (i.e. Laplacian plus potential) Lie
algebraic operator.  One should note that the class of Lie algebraic
operators is closed under gauge transformations.  One is therefore
allowed to construct a Lie algebraic Hamiltonian by gauge-transforming
an arbitrary Lie algebraic $\cH_0$ into self-adjoint form (whenever
such a transformation is possible). Of course, if $W_\lambda$ is an
invariant subspace for $\cH_0$, then $W_\lambda$ multiplied by the
gauge factor will be an invariant subspace for the Hamiltonian, $\cH$.

The above generalized construction involves two complications not
encountered in the 1-dimensional case.  First, every 1-dimensional,
second-order differential operator can be related to a self-adjoint
operator by a gauge transformation.  Essentially, the reason for this
is that all 1-dimensional 1-forms are closed, and of course this is no
longer true in higher dimensions.

Every locally defined second order differential operator can be
expressed in local coordinates as
$$\cH_0=\sum_{ij} g^{ij}\partial_{ij} + \sum_i h^i \partial_i + f.$$
If its symbol, $g^{ij}$ is non-degenerate, then the operator can also
be expressed invariantly as
$$\cH_0=\Delta + \vec{V} + F,$$
where $\Delta$ is the Laplace-Beltrami
operator associated with the pseudo-Rie\-mann\-ian metric structure
$g^{ij}$, and where $\vec{V}$ and $F$ are, respectively, a vector
field and a function. It is not hard to check that $\cH_0$ can be
gauge-transformed into a Schr\"odinger operator if and only if
$\vec{V}$ is a gradient vector field (w.r.t. the $g^{ij}$ metric
structure).  Thus, the condition that $\cH_0$, defined as per
(\ref{laoperator.eqn}), be gauge-equivalent to a Schr\"odinger
operator imposes a severe restriction on the choice of the $C^{ab}$
and $B^a$ coefficients.

Second, in higher dimensions the metric geometry engendered by the
symbol of a second-order differential operator is not in general
Euclidean.  If one imposes the condition of vanishing curvature, then
the choice of admissible $C^{ab}$ coefficients is further restricted.

\subsection{Turbiner's Conjecture}
\label{subsect:tuconj}
The above considerations make clear that Euclidean, Lie algebraic
Hamiltonians are a rather small and particular class of operators.
Surprisingly, in 2 dimensions this class appears to possess an
additional property, namely the corresponding Schr\"odinger equation
can be solved by a separation of variables.  This observation was
first made by A.  Turbiner, who conjectured the following
\cite{turbiner:conj}:
\begin{conjecture}[Turbiner]
  Let $\cH$ be a Lie-algebraic Schr\"odinger operator defined on a
  2-dimensional manifold.  If the symbol of $\cH$ engenders a
  Euclidean geometry, i.e. if the corresponding Gaussian curvature is
  zero, then the spectral equation $\cH\psi=E\psi$ can be solved by a
  separation of variables.
\end{conjecture}
Note that in the above statement and throughout the remainder of the
article, separation of variables will be taken to mean separation
relative to an orthogonal coordinate system \cite{miller:symsep}.

The following example will serve to illustrate the conjecture and will
provide a good reference point for further discussion.  Let $\aal_1$
denote the Lie algebra of affine transformation of the real line.  The
usual realization of this Lie algebra is by vector fields $\partial_u$
and $u\partial_u$.  The present example is based on the related
realization of $\aal_1\oplus\aal_1$ by vector fields on $\reals^2$:
$$T_1=\partial_u,\quad T_2=u\partial_u,\quad T_3=\partial_v,\quad
T_4=v\partial_v.$$
Consider the following Lie algebraic operator:
$$
-\cH_0 = T_1^2 + 2\,\{ T_2+T_4, T_3 \} + a\,(\,2T_2+4T_4) + (4b+2)\,T_3,
$$
where the curly brackets denote the symmetric anti-commutator, and
where $a$ and $b$ are real parameters.  Taking $(u,v)$ as the local
coordinates, the symbol of this operator (up to sign) is given by 
\begin{equation}
  \label{fmetrica.eqn}
  \pmatrix{ 1 & 2u \cr 2u & 4v \cr}.
\end{equation}
Interpreting the above matrix as the contravariant form of a
pseudo-Rie\-mann\-ian metric tensor, $g^{ij}$, one can easily check that
the curvature is zero.  Cartesian coordinates, call them $(x,y)$, are
given by
\begin{equation}
\label{ftrans.eqn}
u=x,\quad v=x^2+y^2
\end{equation}
A straight-forward
calculation shows that $-\cH_0$ can be rewritten as
$$\Delta + 
  \nabla\lp a v + b\log(v-u^2)\rp,$$
where $\Delta$ and $\nabla$ are given  with respect to the metric tensor 
(\ref{fmetrica.eqn}).  
Switching to Cartesian coordinates, one obtains
$$-\cH_0 = \partial_{xx} + \partial_{yy} +
\nabla\lp a (x^2 +y^2)+2b\log|y|\rp.$$
Clearly, $\cH_0$ is gauge-equivalent to a Schr\"odinger operator.
The necessary gauge factor is $\E^\sigma$, where
$$\sigma = 
{a\over 2}
(x^2+y^2) + b\log|y|,$$
and the corresponding
Schr\"odinger operator is
$$
\cH= -\partial_{xx} - \partial_{yy} +2a(1+b) + a^2(x^2+y^2) +
b(b-1)\,y^{-2}.$$
Notice that both $\cH_0$ and the corresponding
Schr\"odinger operator separate in Cartesian coordinates.

Switching to polar coordinates, $r$ and $\theta$, one has
\begin{eqnarray*}
\cH_0 &=& -\partial_{rr} - r^\mone \partial_r  
- r^{-2}\partial_{\theta\theta} -
        \nabla\lp a r^2 + 2b\log(r) + 2b\log|\sin(\theta)|\rp,\\
\cH &=& -\partial_{rr} - r^\mone \partial_r +2a(1+b) + a^2\,r^2
+r^{-2}\left\{-\partial_{\theta\theta} + b(b-1) \sin^{-2}(\theta)\right\}.  
\end{eqnarray*}
It is evident from the above expressions that
$\cH_0$ and $\cH$ also separate in polar coordinates.

\subsection{Summary of results}
\label{sect:sumres}
The purpose of the present article is to show that Turbiner's
conjecture is in general false, but that one can salvage the
conjecture by adding two extra assumptions.  The first of these
assumptions is a somewhat technical compactness condition that serves
to guarantee the completeness of the metric structure induced by the
operator's symbol.  The second assumption is that the generating Lie
algebra acts imprimitively, i.e.  that there exists an invariant
foliation.  Imprimitivity is an indespensible assumption; a
counter-example to the original conjecture based on primitive actions
will be given in Section \ref{sect:counterexample}.  Indeed, the
strength of the imprimitivity assumption is such that in addition to
implying separation of variables, it also implies that the coordinates
of separation will be either Cartesian or polar.

Imprimitivity plays such a decisive role for Turbiner's conjecture
because of the following fact (Corollary \ref{geotraj.cor}): the
perpendicular distribution of the invariant foliation is totally
geodesic.  In other words, a geodesic that is ``launched'' in a
direction perpendicular to the invariant foliation will remain
perpendicular throughout its evolution.  Now on a 2-dimensional
manifold a non-trivial foliation must be 1-dimensional.  Hence the
perpendicular distribution must be 1-dimensional as well, and is
therefore a foliation in its own right.  Hence, if the metric tensor
is Euclidean, and comes from the symbol of a Lie algebraic operator
generated by imprimitive actions, then the leaves of the perpendicular
foliation will be straight lines.

Now there are infinitely many non-isomorphic ways to foliate a small
neighborhood of the Euclidean plane by straight lines.  A foliation of
the full plane is a different matter; one can prove that the leaves of
a global foliation by lines must be mutually parallel.  If one adopts
a slightly more general definition of foliation, then one can prove
that a global foliation must be a pencil of either parallel or
coincident lines (Theorem \ref{strfol.thrm}).  Thus, if one could
somehow globalize the setting of the Lie-algebraic operator to the
full Euclidean plane, then one could prove that the invariant
foliation consists of either parallel lines, or of concentric circles.
The existence of a separating coordinate system --- Cartesian in the
first case, and polar in the second --- readily follows from this
fact.

It turns out that such a globalization is always possible.  The symbol
of a Lie algebraic operator is a tensor that may possess degenerate
points. Indeed, the signature of the tensor can change as one
``crosses'' the locus of degeneracy.  One must therefore throw away
the locus of degeneracy --- it is at most a codimension 1 subvariety
--- and take the domain of the Lie-algebraic Hamiltonian to be one of
those remaining open components where the signature is
positive-definite.  Such a domain is not, in general, isometric to the
Euclidean plane, but with the help of the completeness assumption
alluded to earlier, one can prove that the domain plus a portion of
its boundary is isometric to the Euclidean plane modulo a discrete
group of isometries.

It may help to think of this result as a generalization of the
classical Killing-Hopf theorem \cite{hopf} \cite{stillwell}, which
states that a complete Riemannian manifold of constant curvature is
isometric to the quotient of one the standard space-forms: a sphere,
Euclidean space, or hyperbolic space.  However, the present context
demands a generalization of the notion of pseudo-Riemannian manifold
--- one that permits the metric tensor to have certain well-behaved
singularities.  For lack of better terminology, the name
almost-Riemannian manifold will be used here to refer to manifolds
equipped with such a metric tensor.  It turns out that the symbol of a
Euclidean Lie algebraic operator is an almost Riemannian metric
tensor, and one can therefore use the generalized Killing-Hopf theorem
to obtain the globalization mentioned above.

The proof of the generalized Killing-Hopf Theorem is too long, and too
distinct in scope to be reasonably included in the present article.
The proof is available in the author's PhD dissertation
\cite{milson:thesis}, and will appear in a subsequent publication.
The present article will therefore limit itself to the relevant
definitions and to some hopefully illuminating examples.

The example of Section \ref{subsect:tuconj} conveniently illustrates
the preceding summary.  Regarding the geometric manifestation of
imprimitivity, note that the action of $\aal_1\oplus\aal_1$ is doubly
imprimitive.  The two invariant foliations are $\{u=\mbox{const.}\}$
and $\{v=\mbox{const.}\}$.  The perpendicular foliations are,
respectively, $\{v-u^2=\mbox{const.}\}$ and $\{v/u^2=\mbox{const.}\}$,
or in Cartesian coordinates, $\{y=\mbox{const.}\}$ and $\{y/x
=\mbox{const.}\}$ respectively.  Evidently, both are foliations by
straight lines.

Regarding the notion of an almost-Riemannian manifold, consider the
metric tensor given in (\ref{fmetrica.eqn}). This contravariant metric
tensor has a locus of degeneracy, namely $\{v=u^2\}$. The signature is
positive definite for $v>u^2$ and mixed for $v<u^2$.  The $(u,v)$
plane equipped with this metric tensor turns out to be an instance of
a complete almost-Riemannian manifold.  The domain $\{v\geq u^2\}$ is
isometric to the Euclidean plane modulo the reflection $y\mapsto -y$,
with (\ref{fmetrica.eqn}) giving the corresponding projection.  This
is the sort of phenomenon described by the generalized Killing-Hopf
theorem.

How does separation of variables follow from all this?  Imprimitivity
and the generalized Killing-Hopf theorem allow one to conclude that
the leaves of the invariant foliation are either parallel lines, or
concentric circles.  Indeed, the example of Section
\ref{subsect:tuconj} was chosen to conveniently illustrate both of
these possibilities at the same time.  Now the infinitesimal criterion
for the invariance of a foliation $\{\lambda=\mbox{const.}\}$ is that
the action of a vector field on $\lambda$ must give back a function of
$\lambda$.  It follows immediately that an operator generated by
imprimitively acting vector fields will enjoy the same property.
Therefore, for the example under discussion one is guaranteed that
$\cH_0(x)$ is a function of $x$ and that $\cH_0(r)$, where
$r^2=x^2+y^2$, is a function of $r$.  Since $\cH_0$ is
gauge-equivalent to a Schr\"odinger operator, it must be of the form
$-\Delta + \nabla\sigma$.  Now certainly $\Delta(x)$ is a function of
$x$ and $\Delta(r)$ is a function of $r$, and hence the same is true
if one operates on these two functions with $\nabla\sigma$.  But,
$\nabla\sigma(x) = \partial \sigma/\partial x$ and $\nabla\sigma(r) =
\partial\sigma/\partial r$, where partial derivatives are taken with
respect to Cartesian coordinates in the first instance, and with
respect to polar coordinates in the second.  Hence
$\partial\sigma/\partial x$ must be a function of $x$ and
$\partial\sigma/\partial r$ must be a function of $r$.  Therefore, one
can break up $\sigma$ into a sum of a function of $x$ and a function
of $y$, or into a sum of a function of $r$ and a function of $\theta$.
This makes it quite clear why $\cH_0$ and the corresponding
Schr\"odinger operator separate in Cartesian coordinates, and in polar
coordinates.  The crucial idea of the above argument bears repeating:
if a 2-dimensional, Euclidean, Lie-algebraic operator is imprimitively
generated, then the leaves of the invariant foliation must either be
parallel lines, or concentric circles.

The organization of the remainder of the article is as follows.
Section \ref{sect:prelim} introduces the notation and concepts
necessary for the discussion of Lie-algebraic operators.  Section
\ref{sect:adframe} is concerned with the relevant properties of the
corresponding pseudo-Rie\-mann\-ian geometry.  Section
\ref{sect:armanifold} gives the definition of an almost Riemannian
manifold and Section \ref{sect:arexamples} illustrates the definition
with examples.  Section \ref{strfol.sect} is devoted to the proof of
the result about generalized foliations of the Euclidean plane by
straight lines.  Section \ref{proof.sect} proves the conjecture with
the extra imprimitivity and completeness assumptions, while Section
\ref{sect:counterexample} gives a counter-example to the general form
of Turbiner's conjecture.

%%% Local Variables: 
%%% mode: latex
%%% TeX-master: "paper3"
%%% End: 

\section{The geometry of Lie-algebraic metrics}
\subsection{Preliminaries}
\label{sect:prelim}
Recall from the introduction that a differential operator is called
Lie-alge\-braic if it can be generated by a finite-dimensional Lie
algebra of first-order differential operators.
The purpose of the present section is to describe some properties of
the metric geometry induced by the symbol of a second-order Lie
algebraic operator.

%Recall that the general form of such an operator is
%$$\cH_0 = \sum_{ab} C^{ab} T_a T_b + \sum_a B^a T_a, $$
%where without loss of generality, $C^{ab} = C^{ba}$.
The general form of such an operator, $\cH_0$, is given in
(\ref{laoperator.eqn}).  The symbol, $\sigma(\cH_0)$, is completely
specified by the second-order coefficients, $C^{ab}$.  Indeed,
\begin{equation}
  \label{eq:laopsymbol}
  \sigma(\cH_0) = \sum_{ab} C^{ab} V_a\otimes V_b.
\end{equation}
One endows the domain of $\cH_0$ with the structure of a
pseudo-Riemannian manifold by interpreting $\sigma(\cH_0)$ as the
contravariant form of a metric tensor.  A metric tensor obtained in
this manner will be henceforth referred to as Lie algebraic.

In order for a Lie algebraic metric tensor to be non-degenerate it is
necessary that the vector fields $V_a$ span the tangent space of the
underlying manifold.  This is nothing but the infinitesimal criterion
for transitive action, and therefore the natural setting for a
Lie-algebraic metric is a homogeneous space.  To that end, let $\Gsp$
be a real Lie group and $\Hsp$ a closed subgroup.  Let
$\Msp=\Gsp/\Hsp$ and $\pi:\Gsp\rightarrow\Msp$ denote, respectively,
the homogeneous space of right cosets, and the canonical projection.
For $a\in\gal$ let $a\supl$ and $a\supr$ denote, respectively, the
corresponding left- and right- invariant vector fields on $\Gsp$, and
$\gal\supl$ and $\gal\supr$ the collections of all such.  To avoid any
possible confusion, it should be noted that $\gal\supl$ corresponds to
{\em right} group actions, and $\gal\supr$ to {\em left} ones.  Let
$\ap=\pi_*(a\supl),\; a\in\gal$ denote the realization of $\gal$ by
projected vector fields (i.e. by infinitesimal automorphisms).  It
will also be assumed that $\hal$ does not contain any ideals of
$\gal$.  This will ensure that $a\mapsto\ap$ is a faithful
realization.

The coefficients $C^{ab}$ that specify a Lie-algebraic metric tensor
can be described in a basis-independent manner as an element
$C\in\Sym^2\gal$.  The corresponding Lie algebraic metric tensor is
nothing but $\pi_*(C\supl)$; henceforth it will be denoted simply as $\Cp$.

One must still contend with the fact that $\Cp$ may be a degenerate
tensor.  The projection $\pi:\Gsp\rightarrow \Msp$, naturally induces
a vertical distribution, $\hr$, on $\Gsp$.  Dually, there is the
cotangent sub-bundle of horizontal 1-forms, $\hpr$.  This sub-bundle
is spanned by right-invariant, differential 1-forms $\alpha\supr$,
such that $\alpha\in\gal^*$ annihilates $\hal$.  The extra information
given by $C$ induces a decomposition of the tangent bundle of $\Gsp$,
and allows one to speak of horizontal vectors and vertical 1-forms.
The decomposition is given by
\begin{equation}
\label{tbdecomp.eqn}
\rT\Gsp=\hr\oplus\Cl\hpr.  
\end{equation}
\begin{proposition}
\label{decompfail.prop}
The above decomposition fails precisely in those fibers of
$\pi:\Gsp\rightarrow\Msp$, where the metric tensor, $C\supp$, is
degenerate. In other words, the projection of
the horizontal distribution, $\Cl\hpr$,
spans the tangent space of $\Msp$ precisely
at those points where the metric tensor is non-degenerate.
\end{proposition}

For the remainder of this section fix a $C\in\Sym^2\gal$ such that the
decomposition (\ref{tbdecomp.eqn}) does not fail identically.  In
particular, the rank of $C$ must be greater or equal to
$\dim(\gal)-\dim(\hal)$.  Let $\Msp_0\subset\Msp$ denote the
submanifold where $\Cp$ is non-degenerate, and regard $\Msp_0$ as a
pseudo-Riemannian manifold with metric tensor $\Cp$.

\subsection{The adapted frame}
\label{sect:adframe}
The goal of the present section is to show that the geodesics of a
Lie-algebraic metric can be described as projections of certain vector
fields on the the group $\Gsp$.  The key tool in this section is a
frame of $\rT\Gsp$ adapted to the horizontal-vertical decomposition
given in (\ref{tbdecomp.eqn}).

\begin{definition}
\label{aframe.def}
A vector field on $\Gsp$ of the form $\Cl\alpha\supr$, where
$\alpha\in\hp$,
will be called {\em horizontal}.  
A vector field of the form $a\supr$, where $a\in\hal$, will be called
{\em vertical}.
Consider an adapted basis, $a_1,\ldots,a_r$, of $\gal$,
where the last $r-n$ entries form a basis of $\hal$. Let 
$\alpha^1,\ldots,\alpha^r$ be the adapted dual basis, where the first
$n$ entries span the space of annihilators of $\hal$.  With respect to such a basis,
denote the 
horizontal vector fields by $H^i=C\supl(\alpha^i)\supr$, 
where $i=1,..,n$,
and the vertical vector fields by $V_i=a_i\supr$, where
$i=n+1,\ldots, r$.
As per Proposition \ref{decompfail.prop}, away from the
degenerate fibers the following vector fields
form a basis of $\rT\Gsp$:
$$H^1,\ldots, H^n,V_{n+1},\ldots,V_r.$$
This basis will be called {\em
  an adapted frame of $\Gsp$ relative to $C$}.
\end{definition}
The structure equations of the adapted frame naturally break up into
three types: vertical--vertical, vertical--horizontal, and
horizontal--horizontal.  The first two types are essentially
uninteresting.  They are related to the structure constants, call them 
$c_{ij}^k$, of $\gal$:
$$[a_i,a_j] = \sum_k c_{ij}^k\,a_k,\where i,j,k = 1\ldots r.$$
\begin{proposition}
The vertical--vertical, and the vertical--horizontal structure
constants of the adapted frame are given by
$$[V_i,V_j] = -\sum_k c_{ij}^k V_k,\quad\mbox{and}\quad
[V_i,H^k] = \sum_j c_{ij}^k H^j.$$
\end{proposition}
By contrast,
the horizontal--horizontal type of structure coefficients
are not, in general, constants.
They will be denoted by $A$ and $B$ according to
$$
[H^i,H^j] = \sum_k 2 A^{ij}_k H^k + \sum_l B^{ijl} V_l.
$$
The factor of 2 in the above equation is there to
simplify some later formulas.  These structure coefficients turn out
to play
a fundamental role in the description of the metric geometry of
$\Cp$.

The horizontal vector fields are not, in general, projectable.
Therefore, an expression of the form $\pi_*(H^i)$ is, at best,
a section of
the pullback bundle, $\pi^*(\rT\Msp)$.
To describe the metric geometry, it is necessary to
pull back to $\Gsp$ the covariant derivative  operator, $\nabla$, of
the Levi-Civita connection on $\Msp_0$, 
so that it can operate on such sections.
Speaking geometrically, this is equivalent to pulling back to $\Gsp$ the
parallel transport operators along paths on $\Msp_0$.
\begin{definition}
Let $\gamma$ be a path on $\Gsp$, and let
$X$ be a section of $\rT\Msp_0$ along $\pi\circ\gamma$.
The pullback of the covariant derivative, it will be denoted 
by $\tnabla$,  is defined by
$$\tnabla_{\dot{\gamma}} (X\circ\pi) = 
(\nabla_{\pi_*(\dot{\gamma})} X)\circ\pi.
$$
\end{definition}
Let $X$ be a vector field on $\Gsp$.  
In the sequel it will be convenient to abbreviate $\pi_*(X)$ as
$X\supp$, and $\tnabla X\supp$ simply as $\tnabla X$.
The distinction is important. The reader should keep in
mind that $\tnabla$ operates on sections of $\pi^*(\rT\Msp_0)$, and {\em
not} on sections of $\rT\Gsp$.
The following abbreviation will also be useful:
$$X\cdot Y = X\supp\cdot Y\supp,$$
where $X$, $Y$ are vector fields on $\Gsp$, and the dot on
the right refers to the inner product on $\Msp_0$.
\begin{proposition}
If $X$ is a projectable vector field on $\Gsp$, then
$$\tnabla X = (\nabla X\supp)\circ \pi.$$
\end{proposition}
\begin{proposition}
Let $X$, $Y$ be vector fields, and $f$ a function on $\Gsp$. The
pullback operator satisfies the following analogues of the standard
identities for the covariant derivative:
\begin{eqnarray*}
\tnabla_{fX} Y &=& f\tnabla_X Y, \cr
\tnabla_X (fY) &=& f\tnabla_X Y + X(f)Y\supp.
\end{eqnarray*}
\end{proposition}
\begin{proposition}
\label{ipcompat.prop}
The inner product on $\Msp_0$ is compatible with $\tnabla$.
Furthermore, $\tnabla$ acts in a torsionless manner.
More formally,
let $X$, $Y_1$, $Y_2$ be vector fields on $\Gsp$. Then,
\begin{eqnarray*}
X(Y_1\cdot Y_2) - (\tnabla_X Y_1)\cdot Y_2
- Y_1 \cdot (\tnabla_X Y_2)&=&0,\\
\tnabla_{Y_1} Y_2 - \tnabla_{Y_2} Y_1 -[Y_1,Y_2]\supp &=&0.
\end{eqnarray*}
\end{proposition}
\begin{proof}
This is a direct consequence of the preceding two propositions.
\end{proof}

With these preliminaries out of the way, it is possible to
derive the connection coefficients in terms of the
adapted frame.
\begin{proposition}
\label{vertcd.prop}
The parallel translation of a horizontal vector in a
vertical direction is given by the flow of the corresponding vertical
vector field.  More formally,
$$\tnabla_{V_i} H^j = [V_i,H^j]\supp = \sum_k c_{ik}^j (H^k)\supp.$$
\end{proposition}
\begin{proof}
This is a consequence of the second identity in Proposition
\ref{ipcompat.prop}
\end{proof}

To derive the formula for the horizontal--horizontal connection
coefficients will require a slight technical diversion.  For $i,j,k$
between 1 and $n$, define 
$$T^{ijk} = d(\al^k)\supr(H^i,H^j),$$
where $d$ is the usual exterior derivative.
Note that the resulting expression is skew-symmetric in the first two
indices.
\begin{lemma}
\label{tprops.lemma}
The symbol $T^{ijk}$ satisfies the following identities:
\begin{eqnarray}
\label{ipder.eqn}
H^i(H^j\cdot H^k) &=& T^{ijk} + T^{ikj},\\
\label{hbrak.eqn}
[H^i,H^j]\cdot H^k &=& T^{ijk}-T^{jki} - T^{kij}.
\end{eqnarray}
\end{lemma}
\begin{proof}
Let $a\in\gal$, and $\al\in\hp$.
From
$\cL_{a\supl}(\alr) = 0$, and from the homotopy formula for the Lie
derivative it follows that for every vector field, $X$, on $\Gsp$ 
\beq
\label{adder.eqn}
X(\alr a\supl) = d\alr(X,a\supl).
\eeq
From
$$H^i\cdot H^j
 = \sum_{k,l=1}^r C^{kl}\;\left < (\al^i)\supr ; ( a^k)\supl\right >
\left < (\al^j)\supr; (a^l)\supl\right>.$$
and from (\ref{adder.eqn}) one derives (\ref{ipder.eqn}).
Using the fact that
$$[H^i,H^j]\cdot H^k = \al^k{}\supr ([H^i,H^j]),$$
and the standard formula for the exterior derivative, one obtains
$$[H^i,H^j]\cdot H^k = H^i(H^j\cdot H^k) - H^j(H^i\cdot H^k) - 
d(\al^k)\supr(H^i,H^j).$$
From this and from (\ref{ipder.eqn}), Eq. (\ref{hbrak.eqn}) follows
immediately.  
\end{proof}

\begin{proposition}
\label{hcovder.prop}
The covariant derivative of a horizontal vector-field in a horizontal
direction is given by
$$
\tnabla_{H^i} H^j = {1\over2}\, [H^i,H^j]\supp
 = \sum_k A^{ij}_k (H^k)\supp.
$$
\end{proposition}
\begin{proof}
As a consequence of Proposition \ref{ipcompat.prop},
the analogue of the standard formula for the covariant derivative of
the Levi-Civita 
connection remains valid for non-projectable vector fields.  The formula
in question is 
\begin{eqnarray*}
2\tnabla_{H^i} H^j\cdot H^k &=& 
        H^i(H^j\cdot H^k) + H^j(H^i\cdot H^k) - H^k(H^i\cdot H^j) \\
        &&  - H^i\cdot[H^j,H^k] - H^j\cdot[H^i,H^k] + H^k\cdot[H^i,H^j]
\end{eqnarray*}
Combining the above with the identities in 
Lemma \ref{tprops.lemma} gives
\begin{eqnarray*}
2\,(\alpha^k)\supr(\nabla_{H^i} H^j)
        &=& T^{ijk} - T^{jki} - T^{kij} \\
        &=& (\alpha^k)\supr([H^i,H^j]) \\
\end{eqnarray*}
By fixing $i$, $j$, and varying $k$, one sees
that $\nabla_{H^i} H^j$ must
match  ${1\over2}[H^i,H^j]\supp$.
\end{proof}

The effort that went into the development of the adapted frames
machinery is justified by the next theorem.
\begin{theorem}
  \label{ncgeo.thrm} 
  The integral curves of horizontal vector fields project down to
  geodesics on $\Msp_0$.
\end{theorem}
\begin{proof}
This theorem is a direct consequence of Proposition \ref{hcovder.prop},
which implies that $\tnabla_{H^i} H^i = 0.$
\end{proof}

A word of caution is required at this point.  A horizontal vector
field is not, in general, projectable, and thus does not give
a foliation of $\Msp_0$ by geodesic trajectories.
The theorem merely states that if $\gamma$ is a path in
$\Gsp$ such that $\dot{\gamma} = (\Cl\alr)\circ\gamma$
for some fixed $\alpha\in\hp$,
then the projection $\pi\circ\gamma$ is a geodesic down on $\Msp_0$.

Theorem \ref{ncgeo.thrm} highlights the group-theoretic origins of the
geometry induced by a Lie-algebraic metric tensor.  Indeed, in order
to obtain the geodesics on $\Msp_0$ there is no need to compute the
Christoffel symbols and then to solve the second order geodesic
equation.  The theorem shows that one need only integrate a certain
vector field on the group, and then project the resulting trajectories
down to $\Msp_0$.  Such a computation will be illustrated in Section
\ref{mgeomexample.subsect}.
\subsection{Geometric consequences of imprimitivity}
\label{impga.sect}
In the preceding section it was indicated that the horizontal vector
fields, $H^i$, are  not, in general, projectable.
This is a pity, because, otherwise there would
be a foliation of $\Msp_0$ by geodesic trajectories.  The purpose of the
present section is to discuss a condition that allows for something
almost as good -- the projectability of a portion of the horizontal
distribution.
The condition in question is the imprimitivity
of the group action.

Recall that $\Gsp$ is said to act imprimitively if there exists a
$\Gsp$-invariant foliation on $\Msp$.  A more algebraic criterion is
given by the following \cite{morozov} \cite{golub}.
\begin{proposition}
\label{impcrit.thrm}
Suppose that the isotropy subgroup, $\Hsp$, is connected.
The $\Gsp$-action on $\Msp=\Gsp/\Hsp$
is imprimitive if and only
if $\hal$ is not a maximal subalgebra of $\gal$, i.e.
if and only if there exists a
Lie algebra $\fal$ that is properly intermediate
between $\hal$ and $\gal$.  If an intermediate subalgebra, $\fal$,
does exist, then the invariant distribution is given by
$\pi_*(\fal\supr)$.
\end{proposition}

For the rest of the section suppose that $\Gsp$ acts imprimitively on
$\Msp$.  Fix an intermediate subalgebra, $\fal$.  Let
$\Lambda=\pi_*(\fal\supr)$ be the $\Gsp$-invariant integrable
distribution on $\Msp$, and let $\Lambda^\perp$ denote the
distribution of tangent vectors that are perpendicular to $\Lambda$.
\begin{proposition}
\label{prop:lamperp}
The projection of a horizontal vector field $\Cl\alr,$ where
$\al\in\fal^\perp$ belongs to  $\Lambda^\perp$.  Indeed,
$\Lambda^\perp$ is spanned by these projections.
%The perpendicular distribution, $\Lambda^\perp$ is spanned by the
%projections  of the horizontal vector fields $\Cl\alr$,
%where $\al\in\fal^\perp$.  Furthermore, if a geodesic on $\Msp_0$
%lies in $\Lambda^\perp$ 
%at one point, then it lies in $\Lambda^\perp$ throughout, i.e. it is
%an integral manifold of $\Lambda^\perp$.
\end{proposition}
\begin{proof}
  Fix an $\alpha\in\fal^\perp$, a $p\in \Msp_0$, and consider
  $(\Cl\alr)_q$ at various points, $q\in\Gsp$, in the fiber above $p$.
  From Proposition \ref{impcrit.thrm} one has
  $\pi_*(\fal\supr_q)=\Lambda_p$ at all $q$ above $p$.  Since
  $(\Cl\alr)_q\cdot u$, where $u\in\rT_q \Gsp$, is just $\alr_q(u)$,
  one can infer that the projection of $(\Cl\alr)_q$ is perpendicular
  to $\Lambda_p$ for all $q$ above $p$.  The non-degeneracy assumption
  on $\Cp_p$ implies that $\dim(\Lambda^\perp)$ is equal to the
  codimension of $\fal$ in $\gal$, and hence is equal to
  $\dim(\fal^\perp)$.  Therefore, the projection of
  $\Cl(\fal^\perp)\supr$ spans $\Lambda^\perp$.
\end{proof}

The present context demands the following generalization of the usual
notion of a  totally geodesic submanifold.
\begin{definition}
  A distribution, $\cD$, of a pseudo-Riemannian manifold will be
  called totally geodesic whenever the following is true: if a geodesic
  belongs to $\cD$ at one point then that geodesic is an integral
  manifold of $\cD$.
\end{definition}
It is now possible to state the key result of the present section.
\begin{theorem}
If $\Lambda$ is an invariant foliation, then $\Lambda^\perp$ is
totally geodesic.
\end{theorem}
\begin{proof}
  Let a geodesic, $\gamma$ be given. By Theorem \ref{ncgeo.thrm},
  $\gamma$ is the projection of an integral curve of a horizontal
  vector field, $\Cl\alr,\,\alpha\in\hal^\perp$.  If a geodesic
  belongs to $\Lambda^\perp$ at one point, then $\alpha$ must be in
  $\fal^\perp$.  Consequently, by Proposition \ref{prop:lamperp},
  $\gamma$ belongs to $\Lambda^\perp$ everywhere.
\end{proof}
The following result is needed in the proof of Turbiner's conjecture.
Recall that if $\Lambda^\perp$ is rank 1, i.e if the codimension of
$\fal$ in $\gal$ is equal to $1$, then $\Lambda^\perp$ is integrable.
In particular this occurs if the codimension of $\hal$ in $\gal$ is
$2$, i.e. when $\Msp$ is two-dimensional.
\begin{corollary}
  \label{geotraj.cor}
  If $\rank(\Lambda^\perp)=1$, then the integral curves of
  $\Lambda^\perp$ are geodesic trajectories.  Indeed, in this case the
  geodesics are given by the projection of integral curves of
  $\Cl\alr$ where $\alpha\in\gal^*$ is any non-zero annihilator of
  $\fal$.
\end{corollary}

\subsection{An example}
\label{mgeomexample.subsect}
At this point it will be helpful to illustrate the concepts and formulas 
of the preceding sections with a concrete example.  This
example will be based on the two-dimensional linear representation of
$\GL_2\reals$.  This group is sufficiently ``small''
so as to permit concrete, manageable formulas.

The computations will be based on the group coordinates,
\begin{equation}
\label{glgp.eqn}
\twomat{x&y\\z&w},
\end{equation}
and on the following basis of the lie algebra, $\gl_2\reals$:
$$\begin{array}{llll}
a_1 = \twomat{1&0\\0&0} & a_2=\twomat{0&1\\0&0}  &
a_3 = \twomat{0&0\\1&0} & a_4=\twomat{0&0\\0&1}
\end{array}
$$
The homogeneous space, $\Msp$, is $\reals^2$ minus the origin, and the
projection from the group to $\Msp$ will be the operation of taking the first
row of the coordinate matrix (\ref{glgp.eqn}).
As such, the group coordinates $x$, $y$ also serve
as coordinates on $\Msp$.  This setup induces the following
vector field realization of $\gl_2\reals$   :
$$\begin{array}{llll}
\ap_1 = x\partial_x, & \ap_2 = x\partial_y, &
\ap_3 = y\partial_x, & \ap_4 = y\partial_y
\end{array}
$$
The natural basepoint of $\Msp$ is $x=1$, $y=0$. The isotropy algebra at this
point is spanned by $a_3$ and $a_4$.

Define $C\in\Sym^2\gal$ by
$$C=a_1^2 +  a_4^2 -  a_1\odot a_4 + a_1\odot a_3 + a_2\odot a_4,$$
and consider the corresponding Lie algebraic metric tensor
\begin{eqnarray*}
\label{cp.eqn}
\Cp = \mat{cc}{x^2+2xy & -xy \\
                -xy & 2xy+y^2}.
\end{eqnarray*}
This is a Euclidean metric with Cartesian coordinates $(\xi,\eta)$ given by
\begin{equation}
\label{fcoords.eqn}
x=\E^\xi \sin^2(\eta),\quad y=\E^\xi\cos^2(\eta).
\end{equation}

Since $\GL_2\reals$ is an open subset of the affine space of
two-by-two matrices, one can represent the tangent vectors of the
group by matrices, and conveniently describe vector fields as matrices
with entries that are functions of $x$, $y$, $z$, $w$.  Thus, to get a
left- (respectively right-) invariant vector field one simply left
(respectively right) multiplies a constant matrix by the generic group
element (\ref{glgp.eqn}).  For instance, the right-invariant vector
fields, $a_1\supr,\ldots,a_4\supr$, are represented by
$$
\twomat{x&y\\0&0},\quad
\twomat{z&w\\0&0},\quad
\twomat{0&0\\x&y},\quad
\twomat{0&0\\z&w}.
$$
To describe the horizontal vector fields it is necessary to have an
expression for the
contraction of right-invariant vector fields and left-invariant 1-forms.
To this end one uses the formula
$(\alpha^i{})\supr (a_j)\supl = \Ad^i_j.$
The adjoint representation matrix is
$$
{1\over xw-yz} \left(
\begin{array}{rrrr}
xw      & -xz   & yw    & -yz \\
-xy     & x^2   & -y^2  & xy \\
wz      & -z^2  & w^2   & -wz \\
-yz     & xz    & -yw   & xw
\end{array}
\right).
$$
From this one computes the horizontal vector fields to be
\begin{eqnarray*}
H^1 &=&
\mat{cc}{x&y\\z&w} 
+ {w+z\over xw-yz} \mat{cc}{2xy & - 2xy \\ xw+yz & -xw-yz } 
\\
H^2 &=&
- {x+y\over xw-yz} \mat{cc}{
2xy  & -2xy  \\
xw+yz  & -xw-yz }
\end{eqnarray*}
By Lemma \ref{tprops.lemma}, $H^i(H^i\cdot H^i)= 0$,
and hence
$$
\kappa_1 = H^1\cdot H^1+1  = {2xy(w+z)^2\over(xw-yz)^2},\quad
\kappa_2 = H^2\cdot H^2 =  {2xy(x+y)^2\over(xw-yz)^2},
$$
are constants of motion of $H^1$, $H^2$, respectively.

The next step will be to illustrate Theorem \ref{ncgeo.thrm} by
integrating the horizontal vector fields and showing that their
integral curves project to straight lines on $\Msp$.  The projections
of $H^1$, $H^2$, are represented by the first rows of the respective
matrix representations.  Hence, the projection of $H^1$ is given by
$$
{dx\over dt} = x + \sqrt{2\,xy\kappa_1},\quad 
{dy\over dt} = y - \sqrt{2\,xy\kappa_1},
$$
These equations can be solved by rewriting them as
$$
{d\over dt}(x+y) = x+y,\quad 
{d\over dt}\left(\sqrt{{x\over y}}\,\right)
        = \sqrt{\kappa_1\over2}\lparen{x\over y} + 1\rparen.
$$
The solutions in Cartesian coordinates are
$$
\eta = \sqrt{\kappa_1\over2}\,\xi + \mbox{const}.
$$
The projection of $H^2$ is given by
$$
{dx\over dt} = - \sqrt{2\,xy\kappa_2},\quad 
{dy\over dt} = \sqrt{2\,xy\kappa_2}.
$$
The solutions are simply
$$\xi=\mbox{const}.$$
Thus one sees that the integral curves of $H^1$
and $H^2$ project down to straight lines.

The linear $\GL_2\reals$ actions considered here are imprimitive.
The invariant foliation is given
by the radial lines, $y/x=\mbox{const}$.
In Cartesian coordinates it is given by $\eta=\mbox{const}$.
According to Proposition \ref{impcrit.thrm}
the invariant foliation corresponds to the subalgebra spanned by
$$a_1\supr,\;a_3\supr,\;a_4\supr.$$
The annihilators of this
subalgebra are spanned by $(\alpha^2)\supr$.  Thus, according to
Corollary \ref{geotraj.cor}, $H^2$ must project to a foliation by
straight lines that are perpendicular to the invariant foliation. This
is in accordance with the above calculations, which show that
projections of the integral curves of $H^2$, namely
$\xi=\mbox{const}$, are perpendicular to the invariant foliation,
namely $\eta=\mbox{const}$.

%%% Local Variables: 
%%% mode: latex
%%% TeX-master: "paper3"
%%% End: 

\section{Almost-Riemannian manifolds}
\subsection{Definitions and motivation}
\label{sect:armanifold}
As was mentioned in the introduction, one
cannot directly interpret a type (2,0) Lie algebraic tensor as a
conventional metric tensor.  The difficulty is caused by the presence
of points where the contravariant tensor is degenerate.  The inverse
tensor is singular at such points, and consequently the tangent space
lacks a meaningful inner product.  The present context therefore
requires a suitable generalization of pseudo-Riemannian structure, one
that will embrace the presence of degeneracies in the contravariant
metric tensor, but do so in a way that results in objects that are
reasonably well behaved.

To this end let $\Msp$ be a real, analytic manifold and $g^{ij}$ a
type (2,0) tensor field.  Let $\Ssp_g\subset\Msp$ denote the
corresponding locus of degeneracy; in a chart of local coordinates
this is just the set of points where $\det g^{ij} = 0$.  The
analyticity requirement means that $\Ssp_g$ is either the empty set, a
codimension 1 subvariety, or all of $\Msp$.  Set $\Msp_0=\Msp
\backslash \Ssp_g$ and suppose that $g^{ij}$ is not identically
degenerate.  Consequently, $\Msp_0$ is an open, dense subset of
$\Msp$, the elements of $\Ssp_g$ are boundary points of $\Msp_0$, and
the connected components of $\Msp_0$ are pseudo-Riemannian manifolds.

Let $u$, $v$ be analytic vector fields such that the corresponding
plane section, $u\wedge v$, is non-degenerate on $\Msp_0$, i.e. such
that $|u|^2 |v|^2 - (u\cdot v)^2\neq 0$.  Let $K(u\wedge v)$ denote
the corresponding sectional curvature function.
\begin{definition}
  The pair $(\Msp,g^{ij})$ will be called an almost-Riemannian
  manifold whenever for all $u$, $v$ as above, $K(u\wedge v)$ has
  removable singularities at points of $\Ssp_g$.
\end{definition}
The following two facts follow immediately from the definition.
First, if $\Msp$ is 2-dimensional, then it is enough to suppose that
the Gaussian curvature has removable singularities at $\Ssp_g$.
Second, if sectional curvature is constant on the connected components
of $\Msp_0$, then $\Msp$ is almost-Riemannian.

The degenerate points of an almost-Riemannian manifold naturally break
up into two classes.
\begin{definition}
  A boundary point, $p\in\Ssp_g$, will be called unreachable if all
  smooth curves with $p$ as an endpoint have infinite length.
  Conversely, a boundary point will be called reachable if it can be
  attained by a finite length curve.
\end{definition}
The existence of reachable boundary points necessitates a suitable
generalization of the notion of completeness.  For a geodesic segment
$\gamma:(0,1)\rightarrow\Msp_0$, let $T>1$ be the largest number,
possibly $\infty$, such that $\gamma$ can be extended to a geodesic
with domain $(0,T)$.
\begin{definition}
  Let $\Rsp$ be an open connected component of $\Msp_0$.  One will say
  that $\Msp$ is complete within $\Rsp$ whenever for all geodesic
  segments lying within $\Rsp$, either $T=\infty$, or
  $\lim_{t\rightarrow T}\gamma(t)$ {\rm (}relative to the manifold
  topology{\rm )} is a reachable boundary point of $\Rsp$.
\end{definition}
The following result is very useful in establishing completeness of an
almost-Riemannian manifold.
\begin{proposition}
  \label{prop:compcrit}
  Let $\Rsp$ be as above.  Suppose that the signature of $g^{ij}$ is
  positive definite within $\Rsp$, and that $\Rsp$ is contained in a 
  compact {\rm (}relative to the manifold topology{\rm )} subset of $\Msp$.  Then,
  $\Msp$ is complete within $\Rsp$.
\end{proposition}

The proof of the imprimitive case of Turbiner's conjecture relies on
the following generalization of the Killing-Hopf theorem \cite{hopf}
\cite{stillwell} to the almost-Riemannian context
\cite{milson:thesis}.  Let $\Rsp$ be an open connected component where
the signature of the metric is positive definite, and let
$\overline{\Rsp}$ denote the union of $\Rsp$ and the reachable points
of its boundary.
\begin{theorem}
  \label{thrm:2dkilhopf}
  Assume the following to be true: $\dim \Msp=2$; the Gaussian
  curvature is constant; $\Msp$ is complete within $\Rsp$.  Let $\Fsp$
  denote one of $\reals^2$, $S^2$, or $\Hb^2$ according to the sign of
  the curvature.  Then, there exists an analytic map
  $\Pi:\Fsp\rightarrow\Msp$, such that $\Pi(\Fsp)=\overline{\Rsp}$,
  and such that $g^{ij}$ is the push-forward of the metric tensor on
  $\Fsp$.  Furthermore, $\overline{\Rsp}$ is isometric to the quotient
  $\Fsp/\Gamma$, where $\Gamma$ is the group of isometries $\phi$ such
  that $\Pi=\Pi\circ\phi$.
\end{theorem}

The proof of the above theorem is rather involved, and will be given
in a subsequent publication.  The present article will limit itself to
a number of examples illustrating the salient features of the
almost-Riemannian formalism.  One also expects that the above theorem
continues to hold in dimensions greater than 2, as well as for mixed
signatures.  However, at the present time this must be left as a
conjecture.

\subsection{Examples}
\label{sect:arexamples}
One naturally encounters the notion of an almost-Riemannian manifold
when considering the standard models of constant curvature spaces.
For instance, in the Poincare model of the hyperbolic plane one has
$ds^2 = v^{-2} (du^2+dv^2)$.  Note that the corresponding
contravariant inverse, $v^2(\partial_u{}^2+\partial_v{}^2)$, is
analytic.  Although the convention is to restrict one's attention to
the domain $\{v>0\}$, one can just as well regard the whole $(u,v)$
plane with the given metric tensor as an instance of an
almost-Riemannian manifold.  The $u$-axis need not be discarded.  It
is the locus of degeneracy and consists entirely of unreachable
points.

Next, consider $\Hb^2$ modeled as a hyperboloid in Lorenzian 3-space.
Let $u,v,w$ denote the coordinates on the latter, and consider pushing
forward the hyperboloid's metric structure onto the $(u,v)$ plane via
the obvious projection. Equivalently, one can pull back the covariant
metric tensor via the map 
$$(u,v)\mapsto (u,v,\sqrt{1+u^2+v^2})$$
and invert.  The end result is
the following contravariant tensor:
$$\pmatrix{ 1+ u^2 & uv \cr uv & 1+v^2 }.$$
The determinant of the
above matrix is $1 +u^2+v^2$, and hence the locus of degeneracy is
empty; it consists of ``imaginary points''.  Consequently one can
regard the $(u,v)$ plane with the above contravariant metric tensor as
an instance of an ordinary Riemannian manifold.  

The situation becomes more interesting when one considers the
analogous construction for positive curvature.  Consider the
projection of the unit sphere in Euclidean $(u,v,w)$ space to the
$(u,v)$ plane. Pulling back the Euclidean metric along the map
$$(u,v)\mapsto (u,v,\sqrt{1-u^2-v^2})$$
and inverting, one obtains the
following contravariant tensor:
$$
\pmatrix{ 1- u^2 & -uv \cr -uv & 1-v^2 }.
$$
Now there is a non-empty locus of degeneracy, namely the circle
$u^2+v^2=1$, and one has no choice but to regard the $(u,v)$ plane as
an instance of an almost-Riemannian manifold.  As an illustration of
the generalized Killing-Hopf theorem note that the closed disk
$\{u^2+v^2\leq 1\}$ is isometric to the quotient of the 2-sphere by
the reflection along the $w$ axis.  This example also illustrates the
behaviour of an almost-Riemannian manifold at the reachable locus of
degeneracy. Reachable boundary points are precisely the places where
the rank of the projection from the Euclidean plane to the closed disk
is less than maximal, i.e. the places where the 2-sphere is ``folded'' by
the projection.

Turning next to the case of zero curvature, let $W$ be a finite
reflection group acting on $\reals^n$.  The current example gives a
procedure for realizing $\reals^n/W$ as an almost-Riemannian manifold.
It is well known \cite{humphreys} that the invariants of $W$ are a
polynomial algebra in certain basic invariants.  To obtain an
almost-Riemmanian manifold one simply treats the basic invariants as
if they were coordinates.  Consider, for example, the $k^{\rm th}$
dihedral group acting on the $(x,y)$ plane.  The algebra of invariants
is generated by $u = x^2+y^2$ and by $v=\Re((x+iy)^k)$.  Pushing
forward the Euclidean metric tensor via the map $(x,y)\mapsto (u,v)$
one obtains a tensor whose entries are the 3 possible products of
$\nabla u$ and $\nabla v$ expressed as functions of $u$ and $v$.  An
easy calculation shows that this contravariant tensor is
\begin{equation}
  \label{eq:kfoldmetric}
\pmatrix{ 
  4 u & 2 k v \cr
  2 k v & k^2 u^{k-1} 
}.
\end{equation}
The locus of degeneracy is the cusp curve $v^2=u^k$.  The region
$\{u^k\geq v^2\}$ is isometric to one of the $2k$ closed wedges carved
out by the mirror lines of the reflections in the dihedral group.
Once again one sees that reachable boundary points ``downstairs''
correspond to mirror lines ``upstairs''.

It is also possible to use an almost-Riemannian manifold to realize
the quotient of Euclidean space by an infinite reflection group.  The
approach is the same as in the preceding example; one finds a set of
basic invariants and uses these as coordinates. As an example let $W$
be the group of plane isometries generated by reflections through the
sides of an equilateral triangle.  Equivalently, $W$ is the affine
Weyl group corresponding to the root system of the $\slf_3$ Lie
algebra \cite{humphreys}.  Let $\hal$ denote the diagonal Cartan
subalgebra equipped with the usual Killing inner product, and let
$L_1$, $L_2$, $L_3$ denote the weights corresponding to, respectively,
the first, second, and third diagonal entry of a trace-free diagonal
matrix. Throughout one should keep in mind that $L_3 = -L_1 - L_2$.
Taking $L_1$ and $L_2$ as non-orthogonal coordinates of $\hal$, the
contravariant form of the metric tensor reads
$$
\lp
\begin{array}{rr}
2/3 & -1/3 \\
-1/3 & 2/3  \\
\end{array}
\rp.
$$
Set $z_k = \exp(2\pi \I\,L_k),$ and note that symmetric polynomials
of the $z_k$'s give $W$-invariant functions of $\hal$.  The $z_k$'s
generate the coordinate ring of the corresponding torus of diagonal
unimodular matrices, and it is known that the algebra of invariant
elements of the complexified coordinate ring is generated by the
characters of the two fundamental representations of $\slf_2\cmplx$
\cite{fulton-harris}:
$$
\chi_{{}_1} = z_1+z_2+z_3,\quad \chi_{{}_{1,1}} = z_1 z_2 + z_2 z_3 +
z_1 z_3.
$$
Calculating formally one obtains
\begin{eqnarray*}
\nabla \chi_{{}_1}\cdot\nabla \chi_{{}_1}  &=& 4\pi^2\lp -{2\over3}\,
\chi_{{}_1}^2 + 2\,\chi_{{}_{1,1}}\rp,\\ 
\nabla \chi_{{}_1}\cdot \nabla \chi_{{}_{1,1}}  &=& 4\pi^2\lp - {1\over3}\,
\chi_{{}_1}\chi_{{}_{1,1}}+3\rp,\\
\nabla \chi_{{}_{1,1}}\cdot \nabla \chi_{{}_{1,1}}&=& 4\pi^2\lp -
{2\over3}\,\chi_{{}_{1,1}}^2 + 2\,\chi_{{}_1}\rp.  
\end{eqnarray*}
On the real torus the two characters are complex conjugates, and so
fundamental invariants are given by the real and imaginary parts of
$\chi_{{}_1}$ , call them respectively $u$ and $v$.  The corresponding
contravariant metric tensor in $(u,v)$ coordinates is given by:
\begin{equation}
  \label{eq:deltoidmetric}
{2\pi^2\over3}
\pmatrix{
- 3\,u^2+v^2 +6\,u +9 &
-4\,uv -6\,v \cr
-4\,uv -6\,v &
-3\,v^2+u^2 -6\,u +9
}
\end{equation}
The locus of degeneracy of the above matrix is given by
$$
(u^2+v^2)^2 - 8\,(u^3-3\,uv^2) + 18\,(u^2+v^2)-27=0.
$$
The above is the Cartesian equation of the Euler deltoid
\cite{planecurves}, the curve obtained by rolling a unit circle inside
a circle of radius 3.  For this reason the tensor in
(\ref{eq:deltoidmetric}) will henceforth be referred to as the deltoid
metric.  In Section \ref{sect:counterexample} it will serve as the
basis for a counter-example to Turbiner's conjecture.

Finally, it will be instructive to consider how Proposition
\ref{prop:compcrit} can be used to show the completeness of an
almost-Riemannian manifold. Consider again the metric tensor
(\ref{fmetrica.eqn}) introduced in Section \ref{subsect:tuconj}.  One
proceeds by compactifying $\reals^2$ to
$\reals\Pb^1\times\reals\Pb^1$.  The latter can be covered by the
following four coordinate systems: $(u,v)$, $(u,V)$, $(U,v)$, $(U,V)$,
where $U=u^\mone$ and $V=v^\mone$.  It is straightforward to check
that the tensor (\ref{fmetrica.eqn}) can be continued in a
non-singular fashion to each of these charts.  In the $(U,V)$ chart,
for example, the tensor in question is given by
$$\pmatrix{U^4 & 2\,UV^2 \cr 2\,UV^2 & 4\,V^3}.$$
The locus of
degeneracy of the extended metric tensor is the closed curve
$$\setbrak{v=u^2}\cup\setbrak{U=0}\cup\setbrak{V=0}.$$
It's not hard
to check that the extra points added by the compactification are all
unreachable boundary points, and thus do not meaningfully alter the
underlying geometry.  Since $\reals\Pb^1\times\reals\Pb^1$ is compact,
one can apply Proposition \ref{prop:compcrit} to conclude that the
almost-Riemannian manifold in question is complete in the component
$\{v>u^2\}$.
%%% Local Variables: 
%%% mode: latex
%%% TeX-master: "paper3"
%%% End: 
 
\section{Global foliation of the plane by straight lines}
\label{strfol.sect}
The present section is a discussion of a theorem to the effect that a
foliation (in a suitably general sense) of the Euclidean plane by
straight lines must be either a pencil of parallel lines or a pencil
of coincident lines.  The functions, distributions, and other
mathematical objects in the present section are assumed to be
real-analytic.

Let $\cD$ be a distribution on $\reals^2$ whose rank at any given
point is either 1 or 2.  This means that locally $\cD$ is given by the
kernel of a non-vanishing analytic 1-form.  Analyticity implies that
the points of rank 1 form a dense, open subset of $\reals^2$.  One
should also recall that a rank 1 distribution is automatically
integrable.
\begin{theorem}
  \label{strfol.thrm}
  Suppose the integral manifold of $\cD$ at every rank 1 point is a
  straight line. Then, there exists a system of Cartesian coordinates,
  $(x,y)$ such that $\cD$ contains the kernel of either $dx$ or of
  $-y\,dx +x\,dy$.  In other words, the collection of integral
  manifolds of $\cD$ will contain either a pencil of parallel lines or
  a pencil of coincident lines.
\end{theorem}
\begin{proof}
  Choose Cartesian coordinates $(x,y)$ such that $\cD$ has rank one at
  the origin, and such that the integral line at the origin is
  vertical.  Thus, near the origin $\cD$ is the kernel of a locally
  defined 1-form, $f\,dx+g\,dy$, such that $g(0,y)$ is identically
  zero.  In the eventuality that $g(x,y)$ is identically zero, all
  vertical lines are going to be integral manifolds, and one can
  conclude that the kernel of $dx$ is contained in $\cD$.
  
  Suppose then that $g(x,y)$ is not identically zero, and set
  $$h(x,y)={f(x,y)x + g(x,y)y\over g(x,y)}.$$
  On the open set where
  $g\neq0$, only one straight line can be an integral manifold of
  $\cD$.  The slope of this line is $-f/g$, and $h$ is its
  $y$-intercept.  Hence for every $(x,y)$ such that $g(x,y)\neq0$,
  there will be two integral manifolds of $\cD$ passing through the
  point $(0,h(x,y))$: a vertical line, and the line with slope
  $-f(x,y)/g(x,y)$.  But, $\cD$ must have rank 2 at a point where two
  different integral lines intersect, i.e. $f=g=0$ at such a point.
  On the other hand, since $\cD$ was assumed to be rank 1 at the
  origin, $f(0,y)$ is not identically zero, and hence the zeroes of
  $f(0,y)$ are isolated points. Consequently, $h(x,y)$ is a constant,
  call it $k$, and hence $f(x,y)x+g(x,y)(y-k)$ is identically zero.
  Therefore, $\cD$ contains the kernel of $-y'\,dx + x\,dy'$, where
  $y'=y-k$.
\end{proof}

How is the above theorem related to Turbiner's Conjecture?  Recall
that Corollary \ref{geotraj.cor} implies that, if a Lie-algebraic
metric tensor is Euclidean, then the invariant foliation is
perpendicular to straight lines.  The generalized Killing-Hopf theorem
turns this into a global statement.  Theorem \ref{strfol.thrm} is
needed in the proof of the conjecture, because it allows one to
conclude that the globalization of the invariant foliation is either a
pencil of parallel lines, or of concentric circles.  The needed
argument is assembled in the following corollary. 
\begin{corollary}
  \label{invfol.cor}
  Let $\Lambda$ be a rank 1 distribution on a two-dimensional
  manifold, $\Msp$.  Let $\Pi:\reals^2\rightarrow\Msp$ be a map with
  the following properties: there exist points where the Jacobian has
  rank 2, and near such points $\Pi^*(\Lambda)$ is perpendicular to a
  local foliation by straight lines.  Then, there exist Cartesian
  coordinates $(x,y)$ of $\reals^2$, such that $\Pi^*(\Lambda)$
  contains either the kernel of $dy$ or the kernel of $dr$ where
  $r^2=x^2+y^2$.
\end{corollary}
\begin{proof}
  Let $f\, dx + g\,dy$ be a locally defined analytic 1-form whose
  kernel is $\Pi^*(\Lambda)$.  Let $\cD$ be the distribution that is
  locally specified by the kernel of $-g\,dx + f\, dy$.  Consequently
  at points where the Jacobian of $\Pi$ is non-degenerate $\cD$ is
  rank 1 and its integral manifolds are straight lines.  The set of
  such points is open and dense, and hence $\cD$ satisfies the
  hypotheses of Theorem \ref{strfol.thrm}.  The desired conclusion
  follows immediately.
\end{proof}

%%% Local Variables: 
%%% mode: latex
%%% TeX-master: "paper3"
%%% End: 
\section{The conjecture: proof and counter-example}
\subsection{The imprimitive case}
\label{proof.sect}
With the tools developed in the preceding sections it is possible to
prove a form of Turbiner's Conjecture that incorporates two extra
assumptions.  The first assumption is that the generating Lie-algebra
acts imprimitively.  The second assumption is that the domain of the
operator is a homogeneous space, $\Msp=\Gsp/\Hsp$, that is either
compact, or failing that, can be compactified in a $\Gsp$-compatible
manner.  The imprimitivity assumption is indespensible.  Indeed, the
next section presents a counter-example to the conjecture based on
primitive actions.  The compactness assumption implies completeness,
and is needed in order to apply the generalized Killing-Hopf theorem
(Theorem \ref{thrm:2dkilhopf}).

In light of the fact that the generating Lie algebra consists, in
general, of inhomogeneous first-order operators one needs to say a bit
more about the imprimitivity assumption.  The geometric meaning of
imprimitivity is that there exists a foliation such that the group
actions move one leaf to another.  This geometric description cannot
be applied to a Lie algebra of inhomogeneous first-order operators.
One therefore requires the following generalized notion of
imprimitivity.
\begin{definition}
  A collection of operators $\{T_\alpha\}$ will be said to act
  imprimitively if there exists a foliation $\Lambda$ such that for
  all locally defined functions, $\lambda$ whose leaves are the level
  sets of the foliation, and for all $\alpha$, it is the case that
  $T_\alpha(\lambda)$ and $\lambda$ are functionally dependent.
\end{definition}
Let $\Usp$ be an open subset of a a two-dimensional homogeneous
space $\Msp=\Gsp/\Hsp$, where $\gal$, as usual, denotes the Lie
algebra corresponding to $\Gsp$.  Analyticity is an indespensible
assumption in the present context, and so it is important to recall
that a homogeneous space is automatically endowed with a real-analytic
structure \cite{chevalley}.  Let $\eta:\gal\rightarrow\cfty(\Usp)$
be a linear map such that the operators
$$T_a = \ap+\eta(a),\quad a\in\gal,$$
give a realization of $\gal$ by
first order differential operators on $\Usp$.  
\begin{proposition}
  \label{prop:genimp}
  If the operators $\{T_a: a\in\gal\}$ act imprimitively, then so do
  the operators $\{\ap:a\in\gal\}$, i.e. there exists a
  $\Gsp$-invariant foliation on $\Msp$.
\end{proposition}
\begin{proof}
  Let $\Lambda$ be the invariant foliation demanded by the hypothesis,
  and $\lambda$ a locally defined, non-degenerate function such that
  the level sets of $\lambda$ are the leaves of $\Lambda$.
  Imprimitivity means that $T_a(\lambda)$ is a function of $\lambda$
  for every $a\in\gal$.  Furthermore the same is true for
  $T_a(\lambda^2)$.  Hence
  $$\ap(\lambda)=T_a(\lambda^2)/\lambda-T_a(\lambda)$$
  is also a
  function of $\lambda$.
\end{proof}
Let $\cH_0$ be a
second-order Lie-algebraic operator generated by the $T_a$'s as per
(\ref{laoperator.eqn}). Let $C\in\Sym^2\gal$ denote the corresponding
second order coefficients.
\begin{theorem}
\label{tuconj.thrm}
Suppose the following statements are true:
\begin{list}{}{\itemsep 0pt}
\item[\llap{\rm (i)}]
$\cH_0$ is gauge equivalent to a
Schr\"odinger operator; 
\item[\llap{\rm (ii)}]
$(\Usp,\sigma(\cH_0))$ is isometric to a subset of
  the Euclidean plane;
\item[\llap{\rm (iii)}]
the operators $\{T_a:a\in\gal\}$ act
imprimitively; 
\item[\llap{\rm (iv)}]
$\Msp$ is either compact, or can be compactified in
such a way that the $\Gsp$-action on $\Msp$ extends to a real-analytic
action on the compactification.  
\end{list}
Then, both the eigenvalue equation
$\cH_0\psi = E\psi,$ and the corresponding Schr\"o\-dinger equation
separate in either a Cartesian, or a polar coordinate system.
\end{theorem}
\begin{proof}
  By hypothesis (ii), $(\Msp, \Cp)$ is a zero-curvature,
  almost-Riemannian manifold.  By hypothesis (iv) and by Proposition
  \ref{prop:compcrit}, $\Msp$ is complete within the open connected
  component of $\Msp_0$ containing $\Usp$.  Thus, one can apply
  Theorem \ref{thrm:2dkilhopf} to conclude that there exists a real
  analytic map $\Pi:\reals^2\rightarrow\Msp$ such that $\Usp$ is
  contained in the image, and such that $\Cp$ is equal to the
  push-forward of the Euclidean metric tensor.
  
  Let $\Lambda$ be the $T_a$-invariant foliation demanded by
  hypothesis (iii).  By Proposition \ref{prop:genimp}, $\Lambda$ is
  $\Gsp$-invariant as well.  Hence, by Corollary \ref{geotraj.cor}
  $\Pi^*(\Lambda)$, is locally orthogonal to a foliation by straight
  lines.  Next, one can apply Corollary \ref{invfol.cor} to conclude
  that there exist Cartesian coordinates $(x,y)$ such that
  $\Pi^*(\Lambda)=\setbrak{\lambda=\mbox{const.}}$ where $\lambda$ is
  either $y$ or $x^2+y^2$.
  
  Let $x$ and $y$ also denote the corresponding local coordinates down
  on $\Usp$; more precisely, one is speaking of the restriction of
  $x\circ\Pi^\mone$ and $y\circ\Pi^\mone$ to $\Usp$.  By hypothesis
  (i), there exists a function, $\sigma$, such that
  $$\cH_0=-\Delta + \nabla\sigma+U_0.$$
  The invariance of $\Lambda$
  means that $\cH_0(\lambda)$ is a function of $\lambda$, and of
  course 
  \begin{eqnarray*}
     \Delta(f(y)) &=& f''(y), \\
     \Delta(f(x^2+y^2))&=&4f'(x^2+y^2)+4(x^2+y^2)f''(x^2+y^2).
  \end{eqnarray*}
  Consequently $\Lambda$ is invariant with respect to
  $\nabla\sigma+U_0$.  Next, note that
  $$(\nabla\sigma+U_0)(\lambda^2)-\lambda\,(\nabla\sigma+U_0)(\lambda)
  = \lambda\,\nabla\sigma(\lambda),$$
  and hence
  $\nabla\sigma(\lambda)$ and $U_0$ must both be functions of
  $\lambda$.  Hence, as per the discussion in Section
  \ref{sect:sumres}, in the case that $\lambda=y$, one must have
  $\sigma=\sigma_1(x)+\sigma_2(y)$. Similarly, in the case where
  $\lambda=x^2+y^2$, it must be true that $U_0$ is a function of $r$
  and that $\sigma=\sigma_1(r)+\sigma_2(\theta)$, where $(r,\theta)$
  is the corresponding system of polar coordinates.  In the first of
  the above instances, $\cH_0$ and the corresponding Schr\"odinger
  operator separate in Cartesian coordinates.  In the second case, the
  two operators separate in polar coordinates
\end{proof}

\subsection{Counter-example}
\label{sect:counterexample}
This final section describes a counter-example to Turbiner's
conjecture based on the deltoid metric of Section
\ref{sect:arexamples}.  It should be mentioned that the operator
constructed here belongs the well-known class of exactly-solvable
Hamiltonians described by Olshanetsky and Perelomov in
\cite{olshanetsky-perelomov}. These operators arise as a natural
generalization of the Calogero-Sutherland Hamiltonian
\cite{sutherland} and are indexed by the possible finite root systems.
The counter-example operator is the Olshanetsky-Perelomov Hamiltonian
of trigonometric type corresponding to the $A_2$ root system.

It will not be necessary to recall the details of the
Olshanetsky-Perelomov construction.  What is relevant here is the
Lie-algebraic nature of these operators \cite{turbiner:cs}
\cite{turbiner:g2}.  Consider again the deltoid metric tensor
(\ref{eq:deltoidmetric}), but with the $2\pi^2/3$ factor omitted for
convenience.  Since the entries of the matrix in question are second
degree polynomials, this tensor can be generated by the Lie algebra of
infinitesimal affine transformations of $\reals^2$, or $\aal_2$ for
short.  The generators of $\aal_2$ are:
$$
T_1=\partial_u,\quad 
T_2=\partial_v,\quad
T_3 = u\partial_u,\quad
T_4 = u\partial_v,\quad
T_5 = v\partial_u,\quad
T_6 = v\partial_v.
$$
It is not hard to verify, either directly or by using Theorem
\ref{impcrit.thrm} that these operators do not admit an invariant
foliation, i.e. the above realization of $\aal_2$ is primitive.

The Euclidean Laplacian in $(u,v)$ coordinates is given by
\begin{eqnarray*}
\Delta &=& 9\,T_1^2  + 9\,T_2^2 -3\,T_3^2 + T_4^2 + T_5^2 - 3\,T_6^2   \\
       && \quad +  3\,\{T_1, T_3\} - 6\,\{T_1,T_6\} - 3\,\{T_2, T_5\} 
       -4\,\{T_3, T_6\} \\
       && \quad - 3\,T_1 - T_3 - T_6.
\end{eqnarray*}
Next, consider the operator
$$\cH_0 = -\Delta +k\, \nabla \log\sigma,$$
where 
$$\sigma=(u^2+v^2)^2 - 8\,(u^3-3uv^2) + 18\,(u^2+v^2)-27$$
is, up to a
constant factor, the determinant of the tensor matrix
(\ref{eq:deltoidmetric}), and $k$ is a real parameter.  Note that
$\sigma$ is just the square of $(z_1-z_2)(z_2-z_3)(z_1-z_3)$
whence a
straightforward calculation will show that
$$\nabla\log\sigma= -12\,(T_3 + T_6).$$
Consequently, $\cH_0$ is
Lie-algebraic as well.  Since $\cH_0$ is of the form Laplacian plus
gradient, it is gauge-equivalent to a Schr\"odinger operator.
Specifically, 
$$\E^{-{k\over2}\sigma}\circ\cH_0\circ\E^{{k\over2}\sigma} =
-\Delta+U,$$
where the potential in $(u,v)$ coordinates is given by
$$U = - 12\,k^{2} - 12\,k\,(1 + k)\,(u - \sqrt3\,v - 3)\,(u
+\sqrt3\,v-3)\,(2\,u + 3)\, \sigma^\mone,
$$
and in the affine $(L_1,L_2)$ coordinates by 
\begin{eqnarray}
\label{eq:cepot}
U &=&  - 12\,k^{2} + 3\,k\,(1 + k)\left[
{1\over\sin^2(\pi L_1-\pi L_2)}+ \right. \qquad\qquad\qquad\qquad\qquad\qquad\\
\nonumber && \quad\qquad\qquad\qquad\qquad
 \left . + {1\over\sin^2(2\pi L_1+\pi L_2)}+
{1\over\sin^2(\pi L_1+2\pi L_2)}
\right]
\end{eqnarray}

In order to show that the corresponding Schr\"odinger equation cannot
be solved by separation of variables, it will be necessary to recall a
few facts regarding this matter \cite{miller:symsep}
\cite{koorn:defsep} (see also 2.7 of
\cite{ushveridze}).  There exists precisely four types of orthogonal
coordinate systems that can serve to separate a 2-dimensional
Schr\"odinger equation: they are the Cartesian, polar, parabolic, and
elliptic coordinate systems.  The first two of these do not require
further elaboration.  Parabolic coordinates $(u,v)$ are related to
Cartesian coordinates $(x,y)$ by
$$x=(u^2-v^2)/2,\quad y=uv.$$
Elliptic coordinates, $(\xi,\eta)$, are
related to Cartesian coordinates by 
$$x=\cosh\xi\,\cos\eta,\quad y=\sinh\xi\,\sin\eta.$$
\begin{definition}
  Say that a function of $2$ variables, $f$, separates in one of the
  above four coordinate systems whenever $f$ takes one
  of the following forms:
  \begin{itemize}
  \item Cartesian coordinates: $f_1(x)+f_2(y)$;
  \item polar coordinates: $r^{-2}\,[\,f_1(r)+f_2(\theta)\,]$;
  \item parabolic coordinates: $(u^2+v^2)^\mone\,[\,f_1(u)+f_2(v)\,]$;
  \item elliptic coordinates:
    $(\cosh^2\xi-\cos^2\eta)^\mone\,[\,f_1(\xi)+f_2(\eta)\,].$
  \end{itemize}
\end{definition}

\begin{proposition}
  A $2$-dimensional Schr\"odinger equation
  $$-\Delta\,\psi+U\psi=E\psi,$$
  can be solved by separation of variables
  if and only if the potential, $U$, separates in one of the above
  mentioned coordinate systems.
\end{proposition}

The proof that the potential given in (\ref{eq:cepot}) does not
separate relies on the following two lemmas.
\begin{lemma}
  Let $f=1/\sin^2(ax+by)$ and $D=p(x,y,\partial_x,\partial_y)$, where
  $p$ is a polynomial.  Then $Df = 0$ if and only if $D= D_1 \circ
  (b\,\partial_x-a\,\partial_y)$, for some other linear differential
  operator, $D_1$.  Furthermore, if $D$ is a non-zero polynomial in
  the coordinates and coordinate derivations of either the polar or
  parabolic coordinate systems, or a polynomial of $\cosh\xi$,
  $\sinh\xi$ , $\cos\eta$ , $\sin\eta$ , $\partial_\xi$ ,
  $\partial_\eta$, then $Df\neq0$.
\end{lemma}
Let $f_1$, $f_2$, $f_3$ denote the 3 terms inside the bracketed
subexpression of (\ref{eq:cepot}), i.e. $f_1= 1/\sin^2(\pi L_1 - \pi
L_2)$, etc.
\begin{lemma}
  Let $D$ be a non-zero polynomial in the coordinates and coordinate
  derivations of either the Cartesian, polar, parabolic coordinate
  systems, or a $(\xi,\eta)$ operator of the type described in the
  preceding lemma.  Then, $D(f_1+f_2+f_3)\neq 0 $.
\end{lemma}
\begin{proposition}
The Schr\"odinger equation with the potential given in {\rm
  (\ref{eq:cepot})} cannot be solved by separation of variables.
\end{proposition}
\begin{proof}
  It is necessary to show that the potential, $U$, in question does
  not separate in any of the four coordinate systems mentioned above.
  Since $U$ is a smooth function one can use a second-order derivative
  of mixed partials to test for separability.  In other words $U$
  separates in a given Cartesian coordinate system if and only if
  $\partial_{xy} \, U = 0$; it separates in a given polar coordinate
  system if and only if $\partial_{r\theta} \, (r^2 U) = 0$, etc.
  Thus in each case one has to show that $D(U)=0$, where $D$ is the
  appropriate mixed-partials operator.  All such operators fit the
  hypothesis of the preceding lemma, and therefore $D(U)$ can never be
  zero.
\end{proof}
%%% Local Variables: 
%%% mode: latex
%%% TeX-master: "paper3"
%%% End: 


\begin{thebibliography}{10}

\bibitem{paldus}
B.G. Adams, J.~Cizek, and J.~Paldus.
\newblock Lie algebraic methods and their applications to simple quantum
  systems.
\newblock In A.~Bohm, Y.~Ne'eman, and A.O. Barut, editors, {\em Dynamical
  Groups and Spectrum Generating Algebras}, pages 103--208. World Scientific,
  Singapore, 1988.

\bibitem{grpscat}
Y.~Alhassid, F.~G\"ursey, and F.~Iachello.
\newblock Group theory approach to scattering.
\newblock {\em Ann. Phys.}, 148:346--380, 1983.

\bibitem{dgroups}
A.~Bohm, Y.~Ne'eman, and A.O. Barut, editors.
\newblock {\em Dynamical Groups and Spectrum Generating Algebras}.
\newblock World Scientific, Singapore, 1988.

\bibitem{turbiner:g2}
A.~Capella, Rosenbaum M., and A.~Turbiner.
\newblock Solvability of the {$G_2$} integrable system.
\newblock {\em Preprint solv-int/9707005}, 1997.

\bibitem{chevalley}
C.~Chevalley.
\newblock {\em Theory of {Lie} Groups I}.
\newblock Princeton University Press, 1946.

\bibitem{fulton-harris}
W.~Fulton and J.~Harris.
\newblock {\em Representation Theory}.
\newblock Springer-Verlag, 1991.

\bibitem{golub}
M.~Golubitsky.
\newblock Primitive actions and maximal subgroups of {Lie} groups.
\newblock {\em J. Differential Geometry}, 7:175--191, 1972.

\bibitem{GKO:mdim}
A.~Gonzalez-Lopez, N.~Kamran, and P.~J. Olver.
\newblock New quasi-exactly solvable hamiltonians in two dimensions.
\newblock {\em Commun. Math. Phys.}, 159:503--537, 1994.

\bibitem{hopf}
H.~Hopf.
\newblock Zum \hbox{Clifford-Kleinschen} raumformproblem.
\newblock {\em Math. Ann.}, 95:313--339, 1925.

\bibitem{humphreys}
J.~E. Humphreys.
\newblock {\em Reflection Groups and Coxeter Groups}.
\newblock Cambridge University Press, 1990.

\bibitem{KO:sl2}
N.~Kamran and P.~J. Olver.
\newblock Lie algebras of differential operators and {Lie}-algebraic
  potentials.
\newblock {\em J. Math. Anal. Appl.}, 145:342--356, 1990.

\bibitem{koorn:defsep}
T.~W. Koornwinder.
\newblock A precise definition of separation of variables.
\newblock In {\em Proceedings of the Scheveningen Conferences of Differential
  Equations}, Lecture Notes in Mathematics. Springer-Verlag, 1980.

\bibitem{planecurves}
J.~D. Lawrence.
\newblock {\em A Catalog of Special Plane Curves}.
\newblock Dover, 1972.

\bibitem{magyar}
E.~Magyari.
\newblock Exact quantum-mechanical solutions for anharmonic oscillators.
\newblock {\em Physics Letters A}, 81A(2,3):116--119, 1981.

\bibitem{miller:symsep}
W.~Miller.
\newblock {\em Symmetry and Separation of Variables}, volume~4 of {\em
  Encyclopedia of Mathematics and its Applications}.
\newblock Addison-Wesley, 1977.

\bibitem{milson:thesis}
R.~Milson.
\newblock Multi-dimensional {Lie}-algebraic operators.
\newblock {\em Ph.D. Thesis, McGill University}, 1995.

\bibitem{morozov}
V.~V. Morozov.
\newblock On primitive groups.
\newblock {\em Matematichiskii Sbornik}, 5:355--390, 1939.

\bibitem{olshanetsky-perelomov}
M.~A. Olshanetsky and A.~M. Perelomov.
\newblock Quantum integrable systems related to {Lie} algebras.
\newblock {\em Phsyics Reports}, 94(6):313--404, 1983.

\bibitem{turbiner:cs}
W.~R\"uhl and A.~Turbiner.
\newblock Exact solvability of the {Calogero} and {Sutherland} models.
\newblock {\em Modern Physics Letters A}, 10(29):2213--2221, 1995.

\bibitem{ShTu:mdim}
M.~A. Shifman and A.~V. Turbiner.
\newblock Quantal problems with partial algebraization of the spectrum.
\newblock {\em Commun. Math. Phys}, 126:347--365, 1989.

\bibitem{stillwell}
J.~Stillwell.
\newblock {\em Geometry of Surfaces}.
\newblock Springer-Verlag, 1992.

\bibitem{sutherland}
B.~Sutherland.
\newblock Quantum many-body problem in one dimension.
\newblock {\em Math. Phys.}, 12:246--250, 1971.

\bibitem{turbiner:conj}
A.~V. Turbiner.
\newblock Lie algebras and linear operators with invariant subspaces.
\newblock In {\em Lie Algebras, Cohomology, and New Applications to Quantum
  Mechanics}, volume 160 of {\em Contemporary Mathematics}, pages 263--310.
  American Mathematical Society, 1994.

\bibitem{tusl2}
A.V. Turbiner.
\newblock Quasi-exactly-solvable problems and $\rm sl(2)$ algebra.
\newblock {\em Comm. Math. Phys.}, 118(3):467--474, 1988.

\bibitem{ushveridze}
A.~G. Ushveridze.
\newblock {\em Quasi-exactly Solvable Models in Quantum Mechanics}.
\newblock Inst. of Physics Publ., Bristol, England, 1994.

\bibitem{znojil}
M.~Znojil.
\newblock Sextic-oscillator puzzle and its solution.
\newblock {\em Phys. Rev. D}, 34(4):1224--1227, 1986.

\end{thebibliography}
\end{document}